# Electronic band structure, Fermi surface, and the effect of spin-orbit coupling for tetragonal low-temperature superconductor Bi$_2$Pd from first principles

**I. R. Shein\*,  A. L. Ivanovskii**

*Institute of Solid State Chemistry, Ural Branch of the Russian Academy of Sciences, 620990 Ekaterinburg, Russia*

**Abstract.**
 We have examined theoretically the electronic band structure and Fermi surface (FS) of the tetragonal low-temperature superconductor Bi$_2$Pd. Our main results are that (i) the Pd 4$d$ and Bi 6$p$ states determine the main peculiarities of the multiple-sheet FS topology, thus for this material a complicated superconducting gap structure with different energy gaps on different FS sheets should be assumed; (ii) the effect of spin-orbit coupling is of minor importance for the distributions of the near-Fermi electronic states; and (iii) this phase adopts a 3D-like type owing to directional bonds between the adjacent atomic sheets.

*\* E-mail: shein@ihim.uran.ru (I.R. Shein)*

Bi-based and Bi-containing superconductors (SCs) represent very attractive groups of modern superconducting materials. So, Bi$_2$Sr$_2$Ca$_2$Cu$_3$O$_x$ (so-called Bi-2223, with $T_C$ > 110 K [1]) and related phases such as Bi$_2$Sr$_2$CaCu$_2$O$_8$ (Bi-2212) are among the most known members of the high-$T_C$ cuprate family; LaGaBi$_2$ [2] is a member of an
interesting class of SCs with a quasi-one-dimensional structure; a series of layered Bi-containing 1111-like SCs such as LaNiBiO with $T_C$ ~ 4.2 K, LaCuBiO with $T_C$ ~ 6 K, and BiOCuS with $T_C$ ~ 5.8 K has been discovered and analyzed [3-7]; very recently, a novel group of BiS$_2$-based layered superconductors Bi$_4$O$_4$S$_3$ [8] and LaO$_{1-x}$F$_x$BiS$_2$ [9] was reported.

Recently, attention was paid to the Bi-Pd system, where several superconducting phases were found: monoclinic BiPd with $T_C$ ~ 3.8 K, monoclinic Bi$_2$Pd with $T_C$ ~ 1.7 K, tetragonal Bi$_2$Pd with $T_C$ ~ 5.4 K, and hexagonal Bi$_5$Pd$_3$ with $T_C$ ~ 4 K, see [10]. Among them, BiPd belongs to non-centrosymmetric SCs [11], whereas the newest experimental data [10] concerning physical properties of tetragonal Bi$_2$Pd indicate that this material is a multiple-band/multiple-gap superconductor.

Here we present the results of first-principles calculations of the tetragonal Bi$_2$Pd (which is known also as the β-Bi$_2$Pd phase [10]) to understand the peculiarities of its band structure and Fermi surface topology.

The examined β-Bi$_2$Pd phase adopts [10,12] a tetragonal structure, space group *I*4/*mmm*, # 139, Z = 1. This structure can be schematically described as a stacking of square layers of Bi and Pd in the sequence ...Pd/Bi/Bi/Pd... along the direction *z* as shown in Fig. 1. The atomic positions are Bi: 4*e* (0, 0, $z_{Bi}$) and Pd: 2*a* (0, 0, 0).

Our calculations were carried out by means of the full-potential method with mixed basis APW + lo (FLAPW) implemented in the WIEN2k suite of programs [13]. The generalized gradient approximation (GGA) to exchange-correlation potential in the PBE form [14] was used. The full-lattice optimization including the atomic positions was performed. The self-consistent calculations were considered to be converged when the difference in the total energy of the crystal did not exceed 0.1 mRy and the difference in the total electronic charge did not exceed 0.001*e*, as calculated at consecutive steps.

It may be suspected that a strong spin-orbit coupling (SOC) effect inherent to heavy Bi might be involved to affect the properties of this material. Therefore we will compare our results obtained without SOC with the same, where spin-orbit coupling was included as calculated by using a second-variational treatment [23].

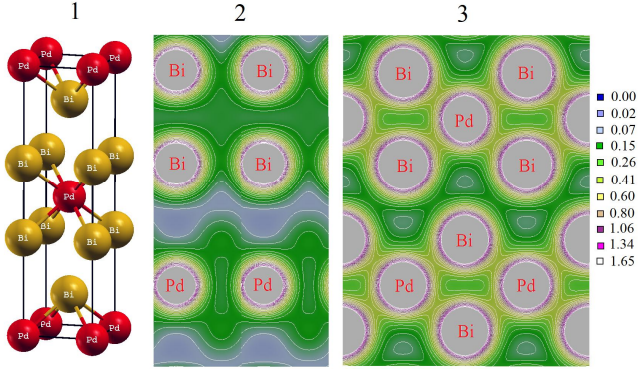

**Fig. 1** (*Color online*) Crystal structure (1) and electronic density maps in: 2 - (100), and 3 - (110) planes for tetragonal β-$Bi_2Pd$.

The structural data are obtained in reasonable agreement with available experiments [10,12], see Table 1.

**Table 1**. The optimized lattice parameters (*a* and *c*, in Å) and internal coordinate ($z_{Bi}$) for tetragonal β-$Bi_2Pd$ in comparison with available experimental data.

| Parameter | *a* | *c* | $z_{Bi}$ |
|---|---|---|---|
| Our data * | 3.4060 / 3.4140 | 13.0115 / 13.0355 | 0.3633 / 0.3616 |
| Experiment ** | 3.362 [1] | 12.983 [1] | 0.363 [1] |
|  | 3.37 [2] | 12.96 [2] | - |

\* as calculated without SOC/with SOC
\*\* [1] – Ref. [12]; [2] – Ref. [10].

Some divergences are related to the well-known overestimation of the lattice parameters within GGA based calculation methods. Besides, the lattice parameters as calculated without SOC and including spin-orbit coupling differ very little: so, $a^{non-SOC} - a^{SOC}$ = 0.008 Å, whereas $c^{non-SOC} - c^{SOC}$ = - 0.024 Å.

Let us discuss the main features of the electronic structure of β-$Bi_2Pd$ as calculated within FLAPW-GGA. In Fig. 2 we depict a band structure for this phase with optimized geometry as calculated along the high-symmetry *k* lines. We see that the manifold of near-Fermi bands demonstrates a complicated character, including quasi-two-dimensional low-dispersive bands, as well as the bands with higher dispersion E(*k*). The Fermi level intersects four bands in all directions between the high-symmetry points of the Brillouin zone, Fig. 2, which form the Fermi surface (FS) of this phase.

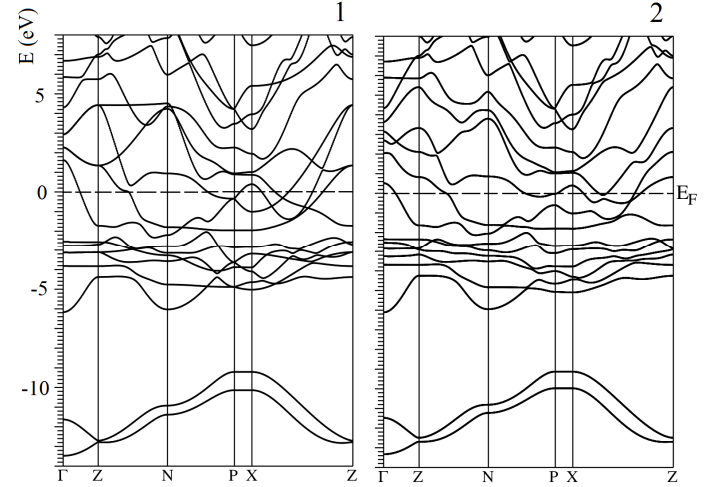

**Fig. 2.** Electronic bands for β-$Bi_2Pd$ as calculated without SOC (1) and with SOC (2).

In turn, the aforementioned features yield an unusual 2D/3D type of multi-sheet FS, which comprises four main hole- and electronic-like pockets, Fig. 3. One of them is a hole 2D-like sheet (along the $k_z$ direction) with the shape of a deformed cylinder. Also, a small closed hole-like pocket is inserted into this cylindrical-like sheet, and thus it is not visible in Fig. 3. A very complicated 3D-like electron pocket covers the aforementioned cylindrical-like pocket and extends to the corners of the Brillouin zone. Finally, one more small closed pocket (along *P-X*) is inserted into this 3D-like electron pocket.

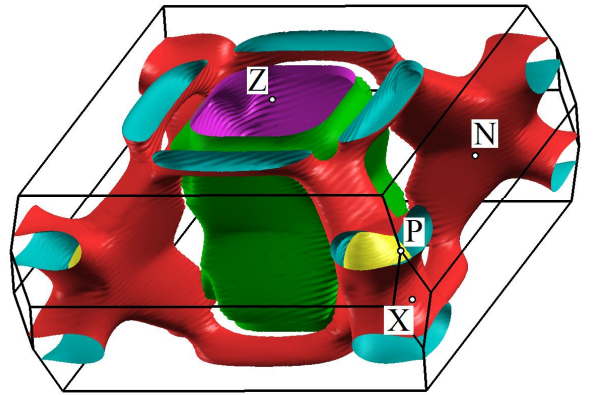

**Fig. 3**. (*Color online*) Fermi surface for β-$Bi_2Pd$.

The presence of multiple Fermi surface suggests that the tetragonal $Bi_2Pd$ should be a multiple-band/multiple-gap superconductor, as was expected from experiments [10].

In Fig. 4 we present densities of states (DOSs) of β-Bi$_2$Pd as calculated without SOC and including spin-orbit coupling. The low-lying DOS peak (from -13.5 eV to -9 eV) belongs to the Bi 6$s$ states. The Pd 4$d$ states are around ~ 3 eV below the E$_F$, whereas the Bi 6$p$ states contribute to the broad spectral interval from -6 eV to E$_F$. We see an overlapping of the valence Pd - Bi states, which form directional bonds between the adjacent Bi/Pd sheets, see below.

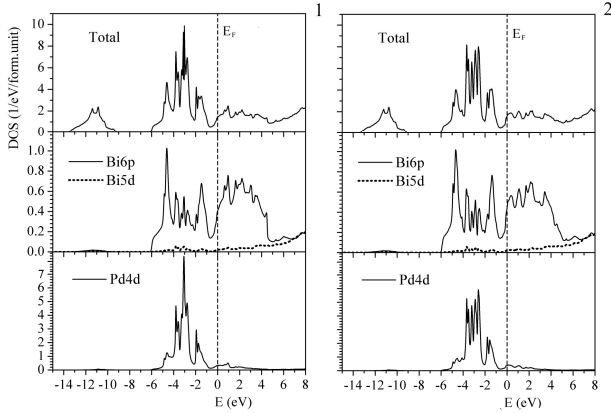

**Fig. 4**. Densities of states for β-Bi$_2$Pd as calculated without SOC (1) and with SOC (2).

The DOS around the Fermi level is of a mixed character and is formed mainly by Pd 4$d$ and Bi 6$p$ states. In turn, the maximal contributions to the DOSs at the Fermi level, N(E$_F$), come from Bi 6$p_{x+y}$, Bi 6$p_z$, and Pd 4$d_{xz+yz}$ orbitals, see Table 2. Therefore both sublattices of β-Bi$_2$Pd are responsible for the metallic properties of this material. In addition, our results demonstrate that the Fermi level is located on the slope of a DOS peak, so the electron doping of β-Bi$_2$Pd increases N(E$_F$), while hole doping will decrease it.

**Table 2** Partial densities of states at the Fermi level (N(E$_F$), in states/eV·atom) for β-Bi$_2$Pd as calculated without SOC and with SOC.

| N(E$_F$) | Bi 6$s$ | Bi 6$p_z$ | Bi 6$p_{x+y}$ | Bi 5$d$ | Pd 5$s$ |
|---|---|---|---|---|---|
| non-SOC | 0.04 | 0.12 | 0.29 | 0.02 | 0.02 |
| SOC | 0.04 | 0.13 | 0.33 | 0.02 | 0.02 |
| N(E$_F$) | Pd 5$p$ | Pd 4$d_{z^2}$ | Pd 4$d_{xy}$ | Pd 4$d_{x^2-y^2}$ | Pd 4$d_{xz+yz}$ |
| non-SOC | 0.04 | 0.03 | 0.08 | 0.03 | 0.19 |
| SOC | 0.04 | 0.04 | 0.10 | 0.04 | 0.20 |

In our calculations with SOC, the value of N$^{tot}$(E$_F$) is only by 11% higher than that without SOC - mainly owing to an insignificant growth of the density of Pd 4$d$ and Bi 6$p$ states, see Table 2. Thus, we conclude that the effect of spin-orbit coupling for β-Bi$_2$Pd is of minor importance for the lattice parameters and for the distribution of the near-Fermi electronic states. On the other hand, certainly, SOC interactions determine the spin-orbit splitting of low-lying core levels. So, according to our calculations, the splitting Δ(Bi 4$d_{5/2}$ - Bi 4$d_{3/2}$) reaches about 2.9 eV - in reasonable agreement with XPS estimations: Δ = 3.1 ± 0.1 eV for crystal Bi and Δ = 2.9 ± 0.1 eV for Bi$_2$Se$_3$ [16].

Finally, the character of bonding in β-Bi$_2$Pd can be well understood from the valence charge density maps, Fig. 1. We see that this phase adopts a 3D-like type owing to the presence of four main types of directional bonds: Bi-Bi (inside square sheets of Bi), Bi-Bi (between the adjacent Bi/Bi sheets), Pd-Pd (inside square sheets of Pd), and Bi-Pd (between the adjacent Bi/Pd sheets). Note that the adjacent Bi/Pd sheets are coupled much more strongly than the adjacent Bi/Bi sheets; besides, the inter-sheet Bi-Pd bonds appear also stronger as compared with Bi-Bi or Pd-Pd bonds inside the corresponding sheets.

In summary, we have probed the electronic band structure and Fermi surface of the tetragonal low-temperature superconductor Bi$_2$Pd. We found that the Pd 4$d$ and Bi 6$p$ states form the near-Fermi bands and determine the main peculiarities of the multiple-sheet FS topology; thus, a complicated superconducting gap structure with different energy gaps on different FS sheets should be assumed for this material. The effect of spin-orbit coupling is of minor importance for the distributions of the near-Fermi electronic states. We found also that for Bi$_2$Pd there are directional bonds (Bi-Bi, Pd-Pd, and Pd-Bi), and this phase adopts a 3D-like type owing to the directional bonds between the adjacent atomic sheets.